# Phase fields in momentum space of photonic crystal slabs*

LI Chuanlin[1], REN Aobo[1,*], WU Jiang[1,2,*]

1. Institute of Fundamental and Frontier Sciences, University of Electronic Science and Technology of China, Chengdu 611731, China
2. Mozi Laboratory, Zhengzhou 450001, China

**Abstract** Optical phase modulation is of great significant importance in fields such as optical communication, information processing, and precision measurement. Compared with real-space modulation, momentum-space phase modulation has distinct advantages: it is free from structural center constraints, supports unlimited mode capacity, and possesses intrinsic topological protection. This inherent flexibility and scalability allow practical applications systems to operate without stringent optical alignment while providing a large number of independent control channels, thereby advancing the development of high-performance, highly integrated optical systems. Photonic crystal slabs, with their open boundary periodicity and capabilities for momentum-space optical field manipulation, have become a crucial platform for research on momentum-space phase fields. Based on polarization orthogonal decomposition and the scattering matrix within temporal coupled-mode theory, this paper systematically elucidates the generation mechanisms of both two-dimensional momentum-space phase fields, including phase vortices, phase gradients, and phase difference, and multidimensional synthetic momentum-space phase fields in photonic crystal slabs, and reviews recent research and application progress in this area. Finally, the development status, advantages, and possible breakthroughs in the field of momentum-space phase fields are summarized and prospects for future work are discussed.

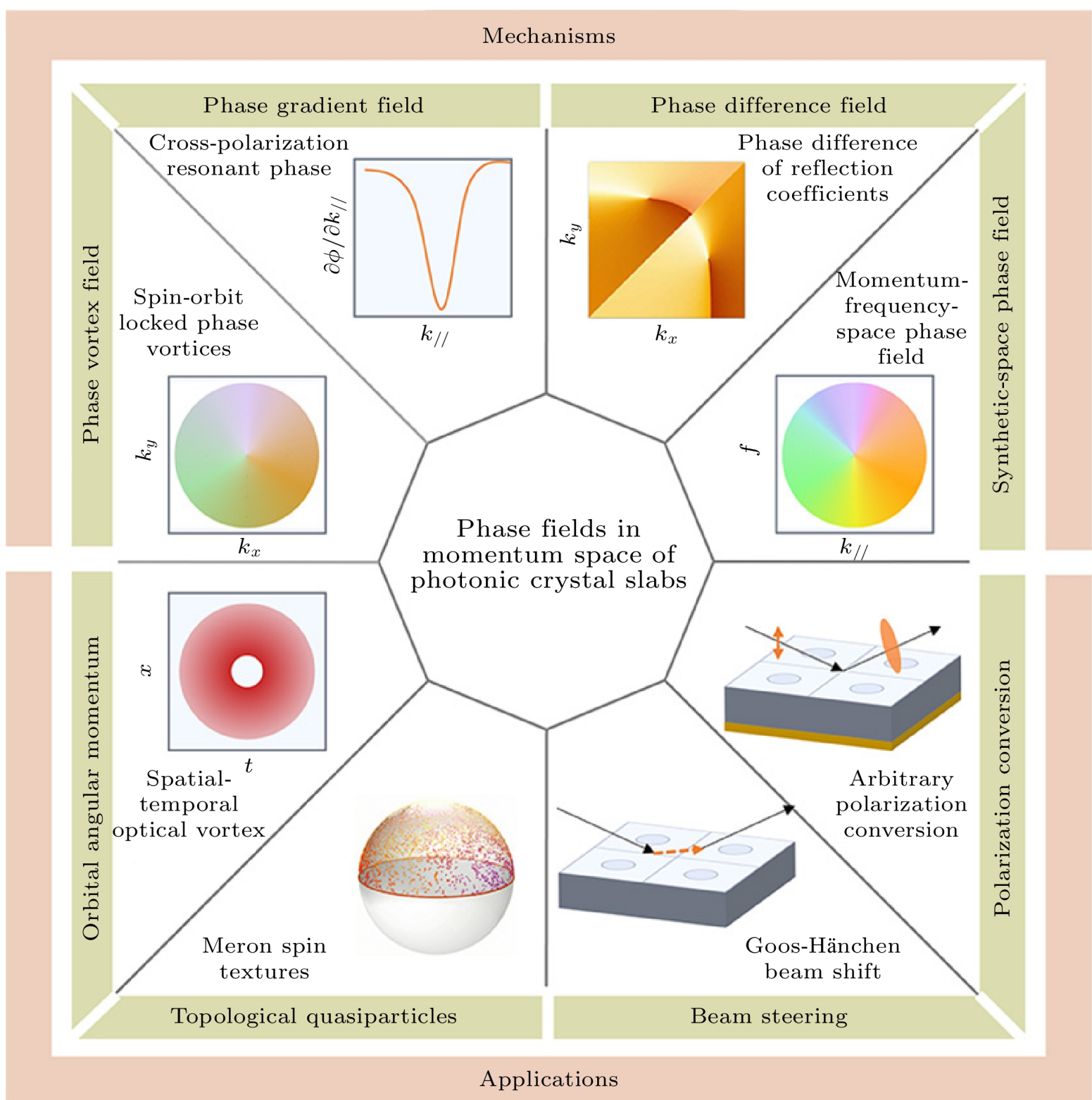

# 1 Introduction

Optical state manipulation represents a critical research direction in modern optics and photonics. Its advancements directly enhance the fundamental research and application performance of optoelectronic devices, including low-power light sources, ultrafast modulators, intelligent detectors, and miniaturized heterogeneous integration systems. Phase, amplitude, wavelength, and orbital angular momentum constitute the fundamental properties of light waves. Among these, phase manipulation holds significant importance in fields such as optical communications and quantum optics[1]. Early explorations of optical phase manipulation began with methods in real space. Classical optical components, such as $\pi/2$ phase converters, spiral phase plates, forked gratings, and spiral zone plates, can generate optical vortices carrying orbital angular momentum by modulating the azimuthal phase of light beams[2]. A 4f pulse shaping system, composed of lenses, diffraction gratings, and spatial light modulators, facilitates the generation of spatiotemporal optical vortices[3,4]. Unlike traditional optical elements that rely on gradual phase accumulation, metasurfaces introduce phase discontinuities at the wavelength scale, significantly reducing the size of optical phase manipulation systems[5-8]. Spatially varying polarization control further enables the realization of geometric phases, known as Pancharatnam-Berry phases. Metasurfaces designed using Pancharatnam-Berry phases can impart distinct orbital angular momenta to incident left- and right-handed circularly polarized light[9]. By manipulating the phase, incident light with other polarization states can also generate orbital angular momentum after passing through polarization-sensitive metasurfaces[10]. Leveraging Pancharatnam-Berry phases induced by spin-orbit interaction, metasurfaces can achieve quantum entanglement between photon spin angular momentum and orbital angular momentum[11]. Due to their structural flexibility and tunable functionality[12-16], metasurfaces have become a pivotal platform for optical phase manipulation research. However, they typically lack topological protection characteristics[17]. Introducing periodicity into spatial structures offers a viable solution to this limitation[18-20].

Photonic crystals are artificial microstructures characterized by periodic distributions of relative permittivity and permeability in space. Similar to Bloch waves formed by electrons in periodic lattice potentials, electromagnetic wave modes in photonic crystals can be described by Bloch functions. This description leads to the formation of photonic band structures with bandgap features[21]. Initially, photonic bandgaps were applied in photonic crystal microcavities to suppress and regulate the redistribution of spontaneous emission, thereby promoting photon localization[22-25]. With the introduction of topological invariant concepts from condensed matter physics, such as the Chern number[26], a series of topological states have been discovered within photonic bandgaps. These include corner states[27-29], edge states[30-35] (such as quantum Hall states, quantum spin Hall states, quantum valley Hall states, and Su-Schrieffer-Heeger states), and Dirac vortex states[36-38]. Furthermore, topological edge states can form through reflections induced by bulk band edge multipole inversions[39]. Although these topological concepts are primarily established within the framework of Hermitian systems, these topological physical effects also demonstrate significant application value in non-Hermitian photonic systems, such as lasers. This is achieved by introducing mechanisms like first-order Bragg diffraction of first-order gratings or zero-order Bragg diffraction of second-order gratings[40-42].

When a photonic crystal possesses finite thickness (i.e., a photonic crystal slab), the system transitions from a Hermitian to a non-Hermitian regime, enabling energy exchange with the external environment[43]. Waveguide modes located above the light cone couple with free space, generating guided resonances[44,45]. These resonant modes radiate outward from the photonic crystal slab as plane waves. They possess definite momentum $\boldsymbol{k} = (k_x, k_y)$ and polarization states, thereby forming polarization fields in momentum space[46-48]:

$$
\begin{aligned}
(c_x, c_y) = & \left(\iint_{\text{unit cell}} \mathrm{d}x\mathrm{d}y\right)^{-1} \\
& \times \iint_{\text{unit cell}} E\left(x, y, z\right) \mathrm{e}^{-\mathrm{j}(k_x x + k_y y)} \mathrm{d}x\mathrm{d}y.
\end{aligned} \tag{1}
$$

Generally, the polarization state of radiation manifests as elliptical polarization, with linear polarization (zero minor axis length) and circular polarization (equal minor and major axis lengths) representing two special cases[49]. The topological invariant in the momentum-space polarization field is characterized by the topological charge. This charge is determined by integrating the gradient of the angle $\alpha(\boldsymbol{k})$ between the major axis of the polarization ellipse and the horizontal direction along a counterclockwise closed loop $L$[50,51]:

$$
q = \frac{1}{2\pi} \oint_L \mathrm{d}\boldsymbol{k} \cdot \nabla_{\boldsymbol{k}} \alpha\left(\boldsymbol{k}\right). \tag{2}
$$

State manipulation in momentum-space polarization fields primarily centers on polarization singularities, including $V$ points (vortex points) and $C$ points (circularly-polarized points)[52,53]. $V$ points carry integer topological charges and are utilized to generate bound states in the continuum and unidirectional guided resonances[54-58]. Bound states in the continuum exhibit extremely high quality factors. They are widely employed to enhance performance in lasers[59-64], quantum light sources[65,66], sensors[67-69], and light-matter interactions[57,70-72]. Unidirectional guided resonance is an intrinsic unidirectional radiative state that requires no external mirrors[73]. It has been applied to realize devices such as high-speed silicon photonic modulators[74] and beam shifters[75]. $C$ points carry half-integer topological charges[49] and enable applications such as high-performance circular dichroism[76-78] and chiral lasers[79,80]. Furthermore, exceptional points also possess half-integer topological charges. At an exceptional point, the complex frequencies and electromagnetic field distributions of modes become degenerate simultaneously[81]. Exceptional points not only facilitate high-sensitivity sensors[82] but also provide a potential bridge connecting band topology and radiative topology[43]. Exceptional points can couple with bound states in the continuum within momentum-space polarization fields[83] and interact with singularities in momentum-space phase fields[84].

Momentum-space phase fields refer to the phase distribution of physical quantities in momentum space (wavevector space). Compared with real-space phase manipulation, momentum-space phase manipulation offers advantages such as eliminating the need for complex optical alignment and providing topological robustness. Based on the dimensionality of manipulation, these fields fall into two categories. The first is the two-dimensional momentum-space phase field, which depends solely on the wavevectors $k_x$-$k_y$[85-87]. The second is the multidimensional synthetic momentum-space phase field, formed by introducing synthetic dimensions such as frequency and geometric parameters. This expansion of dimensions further enriches the degrees of freedom for light field manipulation, enabling possibilities such as the generation of spatiotemporal vortex beams[88,89]. From the perspective of phase distribution morphology, momentum-space phase fields are mainly classified into three types. Phase vortex fields correspond to applications such as orbital angular momentum generation, topological quasiparticle creation, and the photonic spin Hall effect[90,91]. Phase gradient fields enable manipulations such as lateral beam shifts[75,92]. Phase difference fields relate to effects such as perfect polarization conversion and coherent perfect absorption[93,94]. These three types of phase fields collectively constitute the core physical carriers for momentum-space light field manipulation and have attracted widespread attention.

This study elucidates the formation mechanisms of phase fields in two-dimensional and synthetic momentum spaces. The analysis proceeds from polarization orthogonal decomposition and the scattering matrix within the framework of temporal coupled-mode theory. We discuss development trends and frontier issues in this area. On this basis, we review key applications of these phase fields, including orbital angular momentum generation, beam shifting, topological quasiparticle creation, polarization conversion, and coherent manipulation. We then introduce research progress in high-performance active photonic devices. Finally, we summarize and provide an outlook on the development of frontier areas in momentum-space phase fields.

# 2 Two-Dimensional Momentum-Space Phase Fields

## 2.1 Intrinsic Phase Vortex Fields

In dielectric sphere arrays, far-field radiation exhibits a discrete diffraction continuum with definite polarization[95]. This characteristic prevents the revelation of topological properties of bound states in the continuum through polarization distributions in momentum space. To address this issue, researchers employed the quasi-mode expansion method. They constructed the phase distribution field of the quasi-mode coupling strength $W_0$ in momentum space[96]:

$$\theta(k_0, k_z) = \arg(W_0) = \arg[f(k_0, k_z) + \mathrm{i}g(k_0, k_z)], \tag{3}$$

Here, $k_0$ denotes the vacuum wavenumber, and $k_z$ represents the Bloch wave vector. Phase vortices within this phase field, corresponding to the phase winding number or topological charge, can be used to characterize bound states in the continuum[96]:

$$q = \mathrm{sgn}(\frac{\partial f}{\partial k_0}\frac{\partial g}{\partial k_z} - \frac{\partial g}{\partial k_0}\frac{\partial f}{\partial k_z}). \quad (4)$$

Figure 1(b) illustrates the phase distribution field corresponding to the quasi-mode coupling strength $W_0$, with black arrows indicating the vector $\boldsymbol{j} = \nabla\theta$. The center of the phase vortex is located at the intersection of the nodal lines $f(k_0, k_z) = 0$ and $g(k_0, k_z) = 0$. Based on the topological representation of this phase field, one can observe the annihilation and conservation of topological charges. Figure 1(a) presents two bound states in the continuum that carry phase winding numbers of equal magnitude but opposite signs. Further research into the applications of this phase field remains to be explored. In contrast, momentum-space phase vortex fields designed based on far-field polarization configurations have achieved substantial progress in applications such as orbital angular momentum generation.

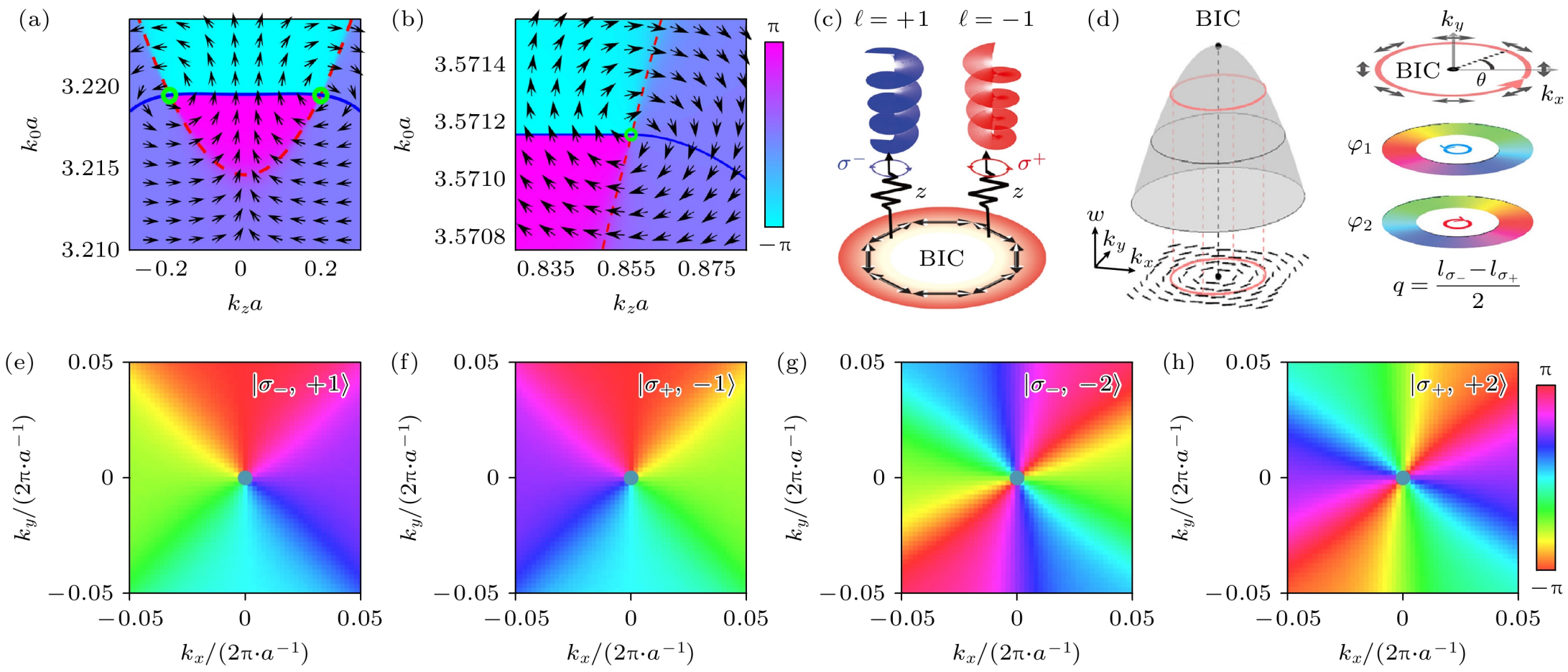

**Fig. 1 Intrinsic phase vortex fields in momentum space: (a), (b) Phase distribution field of the quasi-mode coupling strength $W_0$[96]; (c) schematic illustration of the intrinsic relationship between bound states in the continuum and spatial spiral phase[99]; (d) schematic illustration of the intrinsic relationship between BICs and the winding number of momentum-space phase[85]; (e), (f) low-order phase vortex fields of the circular-polarization component; (g), (h) high-order phase vortex fields of the circular-polarization component[85].**

Orbital angular momentum characterizes the helical phase distribution of light beams and finds significant applications in optical communications, microscopic imaging, and microparticle manipulation. Through spin-orbit locking, the phase vortices carried by spin angular momentum in momentum space can be converted into orbital angular momentum. Near the center of the Brillouin zone in momentum space, the far-field radiation of photonic crystal slabs is predominantly linearly polarized[97]. Any linear polarization state can be expanded using an orthogonal basis composed of left-handed and right-handed circular polarizations. According to Eq. (1), the left-handed circular polarization basis $\sigma_-$ and the right-handed circular polarization basis $\sigma_+$ can be

expressed as

$$|\sigma_+\rangle = \frac{1}{\sqrt{2(|c_x|^2+|c_y|^2)}}(c_x - \mathrm{j}c_y), \quad (5)$$

$$|\sigma_-\rangle = \frac{1}{\sqrt{2(|c_x|^2+|c_y|^2)}}(c_x + \mathrm{j}c_y), \quad (6)$$

A linearly polarized state with a polarization angle of $\varphi$ can be expressed as $|\varphi\rangle = \mathrm{e}^{\mathrm{i}\phi_1}|\sigma_-\rangle + \mathrm{e}^{\mathrm{i}\phi_2}|\sigma_+\rangle$. Here, $\phi_1$ and $\phi_2$ denote the phases of the two circular polarization bases, where $\varphi = (\phi_1 - \phi_2)/2$. Combining this with Eq. (2) yields $q = (l_{\sigma_-} - l_{\sigma_+})/2$. In this expression, $l_{\sigma_-}$ and $l_{\sigma_+}$ represent the winding numbers of phases $\phi_1$ and $\phi_2$ in momentum space, respectively[85]:

$$l_{\sigma_-} = \frac{1}{2\pi}\oint_L \frac{\partial\phi_1}{\partial\theta}\mathrm{d}\theta, \quad l_{\sigma_+} = \frac{1}{2\pi}\oint_L \frac{\partial\phi_2}{\partial\theta}\mathrm{d}\theta. \quad (7)$$

The system exhibits mirror symmetry, satisfying $l_{\sigma_-} = -l_{\sigma_+}$, which implies $q = l_{\sigma_-} = -l_{\sigma_+}$. Figure 1(d) illustrates the intrinsic relationship between bound states in the continuum (BICs) and the phase winding number in momentum space. These phase winding numbers in momentum space are locked to the corresponding orbital angular momentum: the left-handed circularly polarized component carries an optical vortex with a topological charge of $l_{\sigma_-}$, while the right-handed circularly polarized component carries an optical vortex with a topological charge of $l_{\sigma_+}$. Figure 1(c) depicts the intrinsic relationship between BICs and spatial helical phases. Figures 1(e) and (f) show a BIC with a topological charge of $+1$ in a square lattice, corresponding to a phase winding number of $+1$ for the left-handed circularly polarized component and $-1$ for the right-handed circularly polarized component. This spin-locked phase winding rule also applies to higher-order BICs. Figures 1(g) and (h) further demonstrate a BIC with a topological charge of -2[48,98] in a triangular lattice, corresponding to a phase winding number of -2 for the left-handed circularly polarized component and +2 for the right-handed circularly polarized component.

The aforementioned spin-locked helical phases can be equivalently understood from the perspective of the Jones matrix as the effect of applying a circular polarizer to the mode. According to Eqs. (1) and (2), the modes in momentum space can be represented by a two-component Jones matrix[99]:

$$\boldsymbol{E}(\boldsymbol{r},\alpha) = A(\boldsymbol{r})\begin{bmatrix}\cos(q\alpha+\alpha_0)\\ \sin(q\alpha+\alpha_0)\end{bmatrix}, \quad (8)$$

Here, $\boldsymbol{r}$ denotes the position vector of an arbitrary point on the isofrequency contour in momentum space, $A(\boldsymbol{r})$ represents the amplitude, and $\alpha_0$ is the orientation angle of the polarization vector (initial polarization angle) when $\boldsymbol{r}$ aligns with the $+k_x$ direction. The Jones matrices for the left-handed ($\boldsymbol{M}^{\mathrm{L}}$) and right-handed ($\boldsymbol{M}^{\mathrm{R}}$) circular polarizers are given by

$$\boldsymbol{M}^{\mathrm{L}} = \frac{1}{2}\begin{bmatrix} 1 & -\mathrm{i} \\ \mathrm{i} & 1 \end{bmatrix}, \boldsymbol{M}^{\mathrm{R}} = \frac{1}{2}\begin{bmatrix} 1 & \mathrm{i} \\ -\mathrm{i} & 1 \end{bmatrix}. \tag{9}$$

Apply the Jones matrix of the circular polarizer to the Jones matrix of the mode, namely[99]

$$\begin{aligned} \boldsymbol{M}^{\mathrm{L}}\boldsymbol{E}(\boldsymbol{r},\alpha) = & \frac{1}{2}A(r)\mathrm{e}^{-\mathrm{i}(q\alpha+\alpha_0)}\begin{bmatrix} 1 \\ \mathrm{i} \end{bmatrix} \\ = & \frac{1}{2}A(r)\mathrm{e}^{-\mathrm{i}(q\alpha+\alpha_0)}\hat{L}, \end{aligned} \tag{10}$$

$$\begin{aligned} \boldsymbol{M}^{\mathrm{R}}\boldsymbol{E}(\boldsymbol{r},\alpha) = & \frac{1}{2}A(r)\mathrm{e}^{\mathrm{i}(q\alpha+\alpha_0)}\begin{bmatrix} 1 \\ -\mathrm{i} \end{bmatrix} \\ = & \frac{1}{2}A(r)\mathrm{e}^{\mathrm{i}(q\alpha+\alpha_0)}\hat{R}, \end{aligned} \tag{11}$$

Here, $\hat{L}$ and $\hat{R}$ denote the left-handed and right-handed circular polarization bases, respectively. Equations (8) and (9) show that the accumulated phase winding number of the left-handed circularly polarized component in momentum space is $\hat{l^L} = -q$, whereas that of the right-handed circularly polarized component is $\hat{l^R} = q$.

It is worth noting that the aforementioned research primarily faces challenges arising from spin degeneracy. Phase vortices are encoded exclusively within the left-handed and right-handed circularly polarized components, while the overall far-field radiation of the system does not directly carry vortex information. Addressing this challenge will help advance the practical application of spin-locked momentum-space phase vortices.

## 2.2 Phase Vortex Fields Induced by Polarization Conversion

In addition to intrinsic spin-orbit locking (intrinsic spin-orbit interaction), injecting spin angular momentum introduces spin-orbit interaction[100] (extrinsic spin-orbit interaction). Consequently, polarization conversion at the output induces the generation of phase vortex fields. Figure 2(a) illustrates a schematic of this process. In this context, the photonic crystal slab can be modeled as a waveplate. Based on temporal coupled-mode theory[101,102], its transmission matrix under near-normal incidence conditions can be expressed as[86]

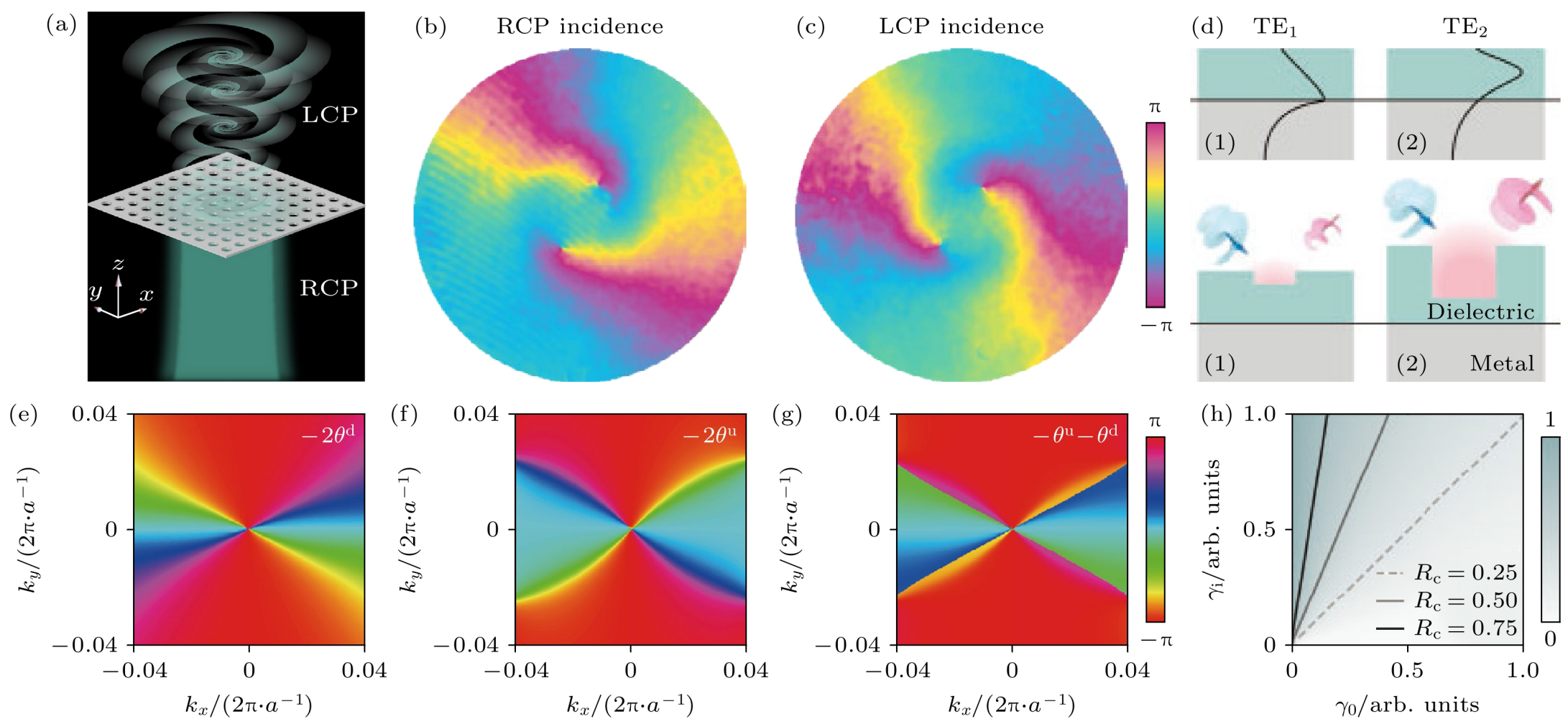

**Fig. 2 Momentum-space phase vortex fields induced by polarization conversion: (a) Schematic of extrinsic spin-orbit interaction[86]; (b), (c) momentum-space phase distribution of transmitted circularly polarized light[86]; (d) mechanism of mode selection for reducing material absorption loss[109]; (e)–(g) transmission- and reflection-type momentum-space phase vortex fields based on Janus bound states in the continuum[87]; (h) relationship between polarization conversion efficiency and material loss $\gamma_i$ as well as radiation loss $\gamma_0$[109].**

$$\begin{bmatrix} -\frac{(d_x^2+d_y^2)(\alpha_z r+t)}{\gamma_0-\mathrm{i}(\omega-\omega_0)}+t & 0 \\ 0 & t \end{bmatrix} = \begin{bmatrix} t_{//}\mathrm{e}^{\mathrm{i}\beta_{//}} & 0 \\ 0 & t_{\perp}\mathrm{e}^{\mathrm{i}\beta_{\perp}} \end{bmatrix}, \tag{12}$$

Here, $(d_x, d_y)$ denotes the coupling coefficients for resonance radiation into external channels, while $r$ and $t$ represent the reflectivity and transmissivity of the direct scattering process, respectively. The term $\alpha_z$ indicates the parity factor associated with out-of-plane mirror symmetry, and $\gamma_0$ represents radiative loss. The right-hand side expresses the general form of the transmission matrix, where $t_{//}$ and $t_{\perp}$ denote the amplitudes of the transmission coefficients parallel and perpendicular to the effective fast axis of the waveplate, respectively, with $\beta_{//}$ and $\beta_{\perp}$ representing the corresponding phases. Light waves with frequencies near the bound states in the continuum are approximately incident normally on the waveplate. The output polarization state can be expressed in the helical basis $|L\rangle$ and $|R\rangle$ as[86]

$$\begin{aligned} |\varphi_{\mathrm{out}}\rangle &= \frac{1}{2}\left(\mathrm{e}^{\mathrm{i}\beta_{//}}t_{//}+\mathrm{e}^{\mathrm{i}\beta_{\perp}}t_{\perp}\right)|\varphi_{\mathrm{in}}\rangle \\ &+ \frac{1}{2}\left(\mathrm{e}^{\mathrm{i}\beta_{//}}t_{//}-\mathrm{e}^{\mathrm{i}\beta_{\perp}}t_{\perp}\right)\mathrm{e}^{-\mathrm{i}2\vartheta}\langle\varphi_{\mathrm{in}}|R\rangle|L\rangle \\ &+ \frac{1}{2}\left(\mathrm{e}^{\mathrm{i}\beta_{//}}t_{//}-\mathrm{e}^{\mathrm{i}\beta_{\perp}}t_{\perp}\right)\mathrm{e}^{\mathrm{i}2\vartheta}\langle\varphi_{\mathrm{in}}|L\rangle|r\rangle, \end{aligned} \tag{13}$$

Here, $|\varphi_{\mathrm{in}}\rangle$ and $|\varphi_{\mathrm{out}}\rangle$ denote the Jones vectors of the input and output light waves. The first term in Eq. (13) represents the phase shift induced by resonance, where the output polarization state remains identical to the input. In contrast, the factor $\mathrm{e}^{\pm\mathrm{i}2\vartheta}$ in the second and third terms corresponds to the introduced geometric phase, specifically the Pancharatnam-Berry phase. When the input light is circularly polarized, the transmitted wave carries the Pancharatnam-Berry phase induced by polarization conversion. Taking right-handed circularly polarized input as an example, the output left-handed circularly polarized wave carries a spiral phase with a topological charge of $l = -2 \times q$. This corresponds to a phase vortex field appearing in the left-handed circularly polarized component within momentum space. Figures 2(b) and 2(c) illustrate the phase distribution in momentum space for transmitted circularly polarized light incident on a bound state in the continuum (BIC) photonic crystal slab with a topological charge of $+1$. These distributions correspond to phase winding numbers of $-2$ and $+2$, respectively.

The preceding discussion addresses conventional BICs, where the topological charges in the upward and downward radiation momentum spaces are equal. The emergence of Janus BICs breaks this symmetry[103-107]. In this case, the transmission Jones matrix must be modified as follows[87]

$$\boldsymbol{J}_{\mathrm{T}}^{\mathrm{f}} = \boldsymbol{J}_{\mathrm{T}}^{\mathrm{b}} = \begin{bmatrix} t + \dfrac{d_{+}^{u} d_{-}^{d}}{\gamma_0 + \mathrm{i}(\omega_0 - \omega)} & \dfrac{d_{+}^{u} d_{+}^{d}}{\gamma_0 + \mathrm{i}(\omega_0 - \omega)} \\ \dfrac{d_{-}^{u} d_{-}^{d}}{\gamma_0 + \mathrm{i}(\omega_0 - \omega)} & t + \dfrac{d_{-}^{u} d_{+}^{d}}{\gamma_0 + \mathrm{i}(\omega_0 - \omega)} \end{bmatrix}, \tag{14}$$

In the equation, the superscripts f and b denote forward $(+z)$ and backward $(-z)$ propagation, respectively. The variables $u$ and $d$ correspond to upward and downward radiation, where $d_{\pm}^{u} = (d_{\mathrm{p}}^{u} \pm \mathrm{i} d_{\mathrm{s}}^{u})/\sqrt{2}$ and $d_{\pm}^{d} = (d_{\mathrm{p}}^{d} \pm \mathrm{i} d_{\mathrm{s}}^{d})/\sqrt{2}$. Here, $d_{\mathrm{p}}$ and $d_{\mathrm{s}}$ represent the coupling coefficients for resonance radiation through the output channel. The subscript p indicates polarization parallel to the plane of incidence, while s indicates polarization perpendicular to the plane of incidence. Equation (15) demonstrates that the transmission Jones matrices for forward and backward propagation are identical, whereas the reflection Jones matrices differ[87]:

$$\boldsymbol{J}_{\mathrm{R}}^{\mathrm{f}} = \begin{bmatrix} r + \frac{d_{+}^{d} d_{-}^{d}}{\gamma_0 + \mathrm{i}(\omega_0 - \omega)} & \frac{d_{+}^{d} d_{+}^{d}}{\gamma_0 + \mathrm{i}(\omega_0 - \omega)} \\ \frac{d_{-}^{d} d_{-}^{d}}{\gamma_0 + \mathrm{i}(\omega_0 - \omega)} & r + \frac{d_{+}^{d} d_{-}^{d}}{\gamma_0 + \mathrm{i}(\omega_0 - \omega)} \end{bmatrix}, \tag{15}$$

$$\boldsymbol{J}_{\mathrm{R}}^{\mathrm{b}} = \begin{bmatrix} r + \frac{d_{+}^{u} d_{-}^{u}}{\gamma_0 + \mathrm{i}(\omega_0 - \omega)} & \frac{d_{+}^{u} d_{+}^{u}}{\gamma_0 + \mathrm{i}(\omega_0 - \omega)} \\ \frac{d_{-}^{u} d_{-}^{u}}{\gamma_0 + \mathrm{i}(\omega_0 - \omega)} & r + \frac{d_{+}^{u} d_{-}^{u}}{\gamma_0 + \mathrm{i}(\omega_0 - \omega)} \end{bmatrix}. \tag{16}$$

Correspondingly, during the conversion from right-handed circularly polarized incidence to left-handed circularly polarized output, the phases of the output light waves

are given by[87]

$$\varphi^{\mathrm{f}}_{\mathrm{R+-}} = -\arg[4\mathrm{i}(\omega_0 - \omega) + 4\gamma_0] - 2\theta^d, \tag{17}$$

$$\varphi^{\mathrm{b}}_{\mathrm{R+-}} = -\arg[4\mathrm{i}(\omega_0 - \omega) + 4\gamma_0] - 2\theta^u, \tag{18}$$

$$\varphi^{\mathrm{f}}_{\mathrm{T+-}} = \varphi^{\mathrm{b}}_{\mathrm{T+-}} = -\arg[4\mathrm{i}(\omega_0 - \omega) + 4\gamma_0] - (\theta^u + \theta^d). \tag{19}$$

The first term in Eqs. (18)-(20) represents the resonant phase, while the second term corresponds to the Pancharatnam-Berry phase dependent on the wave vector. Compared with conventional bound states in the continuum, Janus bound states in the continuum yield richer momentum-space phase vortices. These include the reflected momentum-space phase vortex for forward propagation $l^{\mathrm{f}}_{\mathrm{R+-}} = -2 \times q^d$, the reflected momentum-space phase vortex for backward propagation $l^{\mathrm{b}}_{\mathrm{R+-}} = -2 \times q^u$, and the transmitted momentum-space phase vortex $l^{\mathrm{f}}_{\mathrm{T+-}} = l^{\mathrm{b}}_{\mathrm{T+-}} = -(q^u + q^d)$. Figures 2(e)-(g) illustrate the dependence of momentum-space Pancharatnam-Berry phase vortices on reflection, transmission, and incidence direction when right-handed circularly polarized light is incident on a Janus bound-state-in-the-continuum photonic crystal slab.

Methods that generate momentum-space phase vortices through polarization conversion typically suffer from low efficiency. To address this issue, material loss $\gamma_i$ can be introduced, and only one radiation channel, such as $d_{\mathrm{p}} = 0$, is considered. The efficiency of reflective polarization conversion can then be expressed as[108]

$$R_{\mathrm{c}}(\omega) = \gamma_0^2/[(\omega - \omega_0)^2 + (\gamma_0 + \gamma_{\mathrm{i}})^2], \tag{20}$$

When $\gamma_0 \gg \gamma_{\mathrm{i}}$, high polarization conversion efficiency can be achieved. Figure 2(h) illustrates the relationship between conversion efficiency, material loss $\gamma_{\mathrm{i}}$, and radiation loss $\gamma_0$. Reducing intrinsic material loss and optimizing the design to enhance radiation loss can effectively improve the generation efficiency of momentum-space phase vortices. Figure 2(d) further demonstrates the mechanism for reducing material absorption loss through mode selection. The electric field of the $TM_1$ mode is primarily concentrated near the reflective mirror, resulting in significant loss. In contrast, the electric field of the $TE_2$ mode is weakly distributed near the mirror, thereby substantially reducing material absorption loss. Regulating radiation loss relies not only on mode selection but also on more refined structural design. Future research should further explore methods to enhance the generation efficiency of transmissive momentum-space phase vortices, particularly under conditions involving multiple radiation channels.

## 2.3 Phase Gradient Field

When linearly polarized light is incident on the photonic crystal slab equivalent waveplate, the Pancharatnam-Berry phase generated through polarization conversion manifests as a gradient field rather than a vortex field in momentum space. Taking linearly polarized incidence of $|\pm 45°\rangle$ as an example, the transmission matrix can be expressed as[110]

$$\begin{bmatrix} t-t_{\mathrm{a}} & 0 \\ 0 & t-t_{\mathrm{b}} \end{bmatrix} - \frac{1}{2}(t_{\mathrm{a}}+t_{\mathrm{b}})\frac{\sqrt{S_1^2+S_3^2}}{S_0} \times \begin{bmatrix} 0 & \mathrm{e}^{\mathrm{i}(2\Sigma-\pi/2)} \\ \mathrm{e}^{-\mathrm{i}(2\Sigma-\pi/2)} & 0 \end{bmatrix}, \tag{21}$$

where

$$t_{\mathrm{a}} = \frac{1}{2}\frac{(t+\alpha_z r)(S_0-S_2)}{\gamma_0-\mathrm{i}(\omega-\omega_0)}$$
$$t_{\mathrm{b}} = \frac{1}{2}\frac{(t+\alpha_z r)(S_0+S_2)}{\gamma_0-\mathrm{i}(\omega-\omega_0)}$$
$$\Sigma = \frac{1}{2}\arg(S_3+\mathrm{i}S_1)$$

Here, $S_i$ denotes the Stokes parameters, defined as $S_0=|d_x|^2+|d_y|^2$, $S_1=|d_x|^2-|d_y|^2$, $S_2=2\mathrm{Re}(d_x d_y^*)$, and $S_3=2\mathrm{Im}(d_x d_y^*)$. The first term in Eq. (21) represents co-polarized transmission, whereas the second term describes polarization-conversion transmission. This latter term comprises three components: the resonance parameter $(t_a+t_b)$, the polarization coupling efficiency $\sqrt{S_1^2+S_3^2}/S_0$, and the geometric phase factor matrix. Specifically, the introduced Pancharatnam-Berry phase factor is given by $\Delta\phi_{|\pm 45°\rangle} = \mp[\arg(S_3+\mathrm{i}S_1)-\pi/2]$. We apply a low-order polynomial approximation to $d_x$ and $d_y$ and introduce a perturbation $v$ to $d_x$, such that $d_x = v+\mathrm{i}\mu k_y$ and $d_y = -\mathrm{i}\mu k_x$. Here, $\mu$ represents the polynomial expansion coefficient. When light polarized in the $|+45°\rangle$ state converts to the $|-45°\rangle$ state, the corresponding Pancharatnam-Berry phase is $\Delta\phi_{|+45°\rangle} = -\arg\{4k_x\kappa+\mathrm{i}[4-(k_x^2-k_y^2)\kappa^2]\}+\pi/2$. The phase gradient in momentum space can be expressed as[110]

$$\frac{\partial\left(\Delta\phi_{|+45°\rangle}\right)}{\partial k_x} = \frac{4\kappa\left[4+\left(k_x^2+k_y^2\right)\kappa^2\right]}{16+8\left(k_x^2+k_y^2\right)\kappa^2+\left(k_x^2-k_y^2\right)^2\kappa^4}. \tag{22}$$

It is evident that the maximum phase gradient occurs at the center of the Brillouin zone, where $(k_x, k_y) = (0,0)$. Consider a $Si_3N_4$ photonic crystal slab that breaks in-plane inversion symmetry while preserving in-plane mirror symmetry. When illuminated by linearly polarized light with Stokes parameters $S_2/S_0 = \pm 1$, the Pancharatnam-Berry phase distribution in momentum space appears as shown in Figures 3(a) and 3(b).

Furthermore, plotting the data marked by the orange and gray lines in Figures 3(a) and 3(b) into Figures 3(e) and 3(f) reveals a significant phase gradient along the $k_x$ direction, whereas the variation along the $k_y$ direction remains negligible.

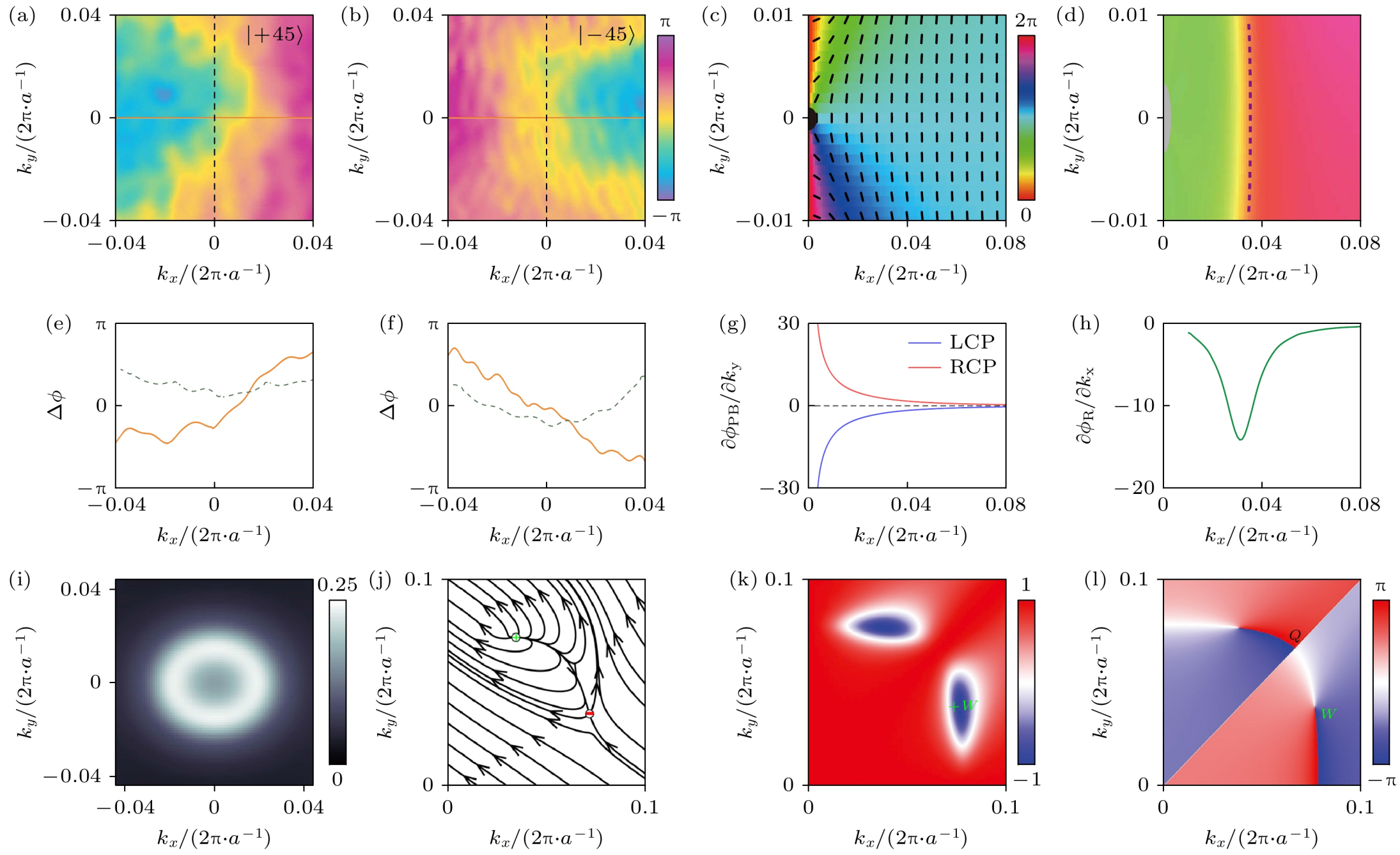

**Fig. 3 Phase gradient fields and phase difference fields in momentum space: (a), (b) Pancharatnam–Berry phase distribution in momentum space under broken inversion symmetry; data marked by orange and gray lines are plotted in (e), (f), respectively[110]; (c), (d) distributions of the polarization-conversion-induced Pancharatnam–Berry phase and the cross-polarization resonant phase in momentum space; the phase gradients along the wavevector directions are shown in (g), (h)[111]; (i) efficiency of momentum-space phase gradient generation via polarization conversion[110]; (j) complex vector field $[\mathrm{Re}(R_{ss}), \mathrm{Im}(R_{ss})]$ in momentum space[93]; (k) distribution of the Stokes parameter $S_1 = |R_{ss}|^2 - |R_{ps}|^2$ in momentum space; (l) phase vortex in the phase difference field $\phi_d(k_x, k_y)$[112].**

The Pancharatnam-Berry phase obtained through polarization conversion is distributed only along a specific direction in momentum space. Studies have shown that when circularly polarized light is incident on a photonic crystal slab supporting bound states in the continuum, it introduces not only a Pancharatnam-Berry phase vortex and its phase gradient along the $k_y$ direction, $\frac{\partial \phi_{PB}}{\partial k_y}$, but also a cross-polarized resonant phase due to nonlocal resonant effects. The gradient of this phase along the $k_x$ direction is[111]

$$\frac{\partial \phi_R}{\partial k_x} = \frac{2\alpha_{v_g}\gamma_0}{4\gamma_0^2 + (|k_{//}| - k_0)^2}, \qquad (23)$$

Here, $k_0$ denotes the wave vector of the guided-mode resonance along the $k_x$ direction,

and $\alpha_{v_g}$ represents the group velocity factor: $\alpha_{v_g} = 1$ corresponds to positive group velocity, whereas $\alpha_{v_g} = -1$ corresponds to negative group velocity. Consider a square-lattice $Si_3N_4$ photonic crystal slab as an example. This structure supports a symmetry-protected bound state in the continuum (BIC) with a topological charge of $-1$. The band hosting this BIC exhibits $\alpha_{v_g} = 1$ near the $\Gamma$ point. When right-handed circularly polarized light with a wavelength near that of the BIC is incident, the Pancharatnam-Berry phase distribution induced by polarization conversion appears as shown in Figure 3(c). This distribution generates a negative phase gradient along the $k_y$ direction (Figure 3(g)). Meanwhile, the distribution of the cross-polarized resonant phase in momentum space is shown in Figure 3(d). It produces a negative phase gradient along the $k_x$ direction (Figure 3(h)). By combining the Pancharatnam-Berry phase gradient with the cross-polarized resonant phase gradient, one can achieve phase gradients in arbitrary directions within momentum space.

Figure 3(i) illustrates the efficiency of generating momentum-space phase gradients via polarization conversion. Improving this efficiency remains a key bottleneck for practical applications. Furthermore, effectively increasing the phase gradient constitutes a critical challenge that requires resolution in this field.

## 2.4 Phase Difference Field

The phase field in momentum space can describe the phase characteristics of a single physical quantity. It can also represent the sum or difference of the phases of two physical quantities. In most physical systems, the value of the absolute phase depends on the choice of the observation point. Consequently, the sum of phases often lacks clear physical significance. In contrast, the phase difference between physical quantities (i.e., relative phase) typically plays a more fundamental role in research. For instance, in double-slit interference, the propagation phase difference directly determines the contrast of the interference fringes. Based on this principle, the following discussion focuses primarily on the phase difference field in momentum space.

For a specific polarization state $[E_s E_p]^T$, the corresponding Stokes parameters are: $S_0 = |E_s|^2 + |E_p|^2$, $S_1 = |E_s|^2 - |E_p|^2$, $S_2 = 2|E_s E_p|\cos\phi_d$, and $S_3 = 2|E_s E_p|\sin\phi_d$. Here:

$$\phi_d = \arg(E_p) - \arg(E_s), \tag{24}$$

represents the phase difference between the s-polarized and p-polarized components. When s-polarized light is incident on a photonic crystal slab and the reflected light is entirely p-polarized, complete polarization conversion occurs, indicated by $|R_{ss}| = 0$. Figure 3(k) illustrates the distribution of the Stokes parameter $S_1 = |R_{ss}|^2 - |R_{ps}|^2$ in momentum space. This point of complete polarization conversion corresponds not only to a saddle point in the complex vector field $[\mathrm{Re}(R_{\mathrm{ss}}), \mathrm{Im}(R_{\mathrm{ss}})]$ formed by the real and imaginary parts of the reflection coefficient (Figure 3(j))[93], but also to a phase vortex in the phase difference field $\phi_d(k_x, k_y)$[112]. As shown in Figure 3(l), the accumulated phase around the complete polarization conversion point $W$ is $2\pi$. This phase structure provides the foundation for achieving arbitrary polarization conversion. Furthermore, for photonic crystal slabs with broken out-of-plane mirror symmetry, the correlation between upward and downward radiation is broken. The phase difference between them can be expressed as[94]

$$\phi_{\mathrm{r}} = \arg(c_{\mathrm{up}}) - \arg(c_{\mathrm{down}}), \tag{25}$$

Here, $c_{\mathrm{up}}$ and $c_{\mathrm{down}}$ represent the far-field polarization complex amplitudes or complex energy flux amplitudes in the upward and downward directions, respectively. Manipulating this phase difference field provides a pathway to achieve effects such as coherent perfect absorption. However, in-depth research on the radiation phase difference field $\phi_{\mathrm{r}}(k_x, k_y)$ in momentum space remains to be conducted.

# 3 Three-dimensional synthetic momentum space phase fields

## 3.1 Phase fields in momentum-frequency space

In recent years, synthetic dimensions have attracted widespread attention[113-119]. They enable the exploration of physical phenomena in higher-dimensional spaces by overcoming the limitations imposed by the apparent geometric dimensions of photonic structures. This section primarily reviews phase fields within synthetic spaces.

Section 2.2 introduced how incident circularly polarized light on a single-layer photonic crystal slab supporting bound states in the continuum (BICs) can introduce Pancharatnam-Berry phase vortices in momentum space. These correspond to spiral phases in real space and longitudinal orbital angular momentum. However, this phase is confined to momentum space and does not appear in the momentum-frequency space. This limitation hinders the application of polarization-conversion-induced Pancharatnam-Berry phases in generating spatiotemporal vortices and transverse orbital angular momentum. Studies indicate that stacking two photonic crystal slabs (Figure 4(a)) introduces interlayer coupling effects. This effect is closely related to the interlayer gap $h$ and the in-plane relative displacement $\boldsymbol{d}$[88]. When the in-plane horizontal positions of the two slabs are perfectly aligned ($\boldsymbol{d} = 0$), the structure supports multiple BICs, similar to a single-layer slab. Upon incidence of circularly polarized light, phase vortices form in momentum space (Figure 4(b)). However, only a parabolic distribution appears in the frequency-momentum space (Figure 4(c)). Once an in-plane horizontal offset is introduced ($\boldsymbol{d} \neq 0$), the out-of-plane mirror symmetry is broken. This not only leads to radiation asymmetry but also induces clear phase vortices in the momentum-frequency space (Figure 4(d)). Consequently, this provides new possibilities for realizing spatiotemporal vortices and controlling transverse orbital angular momentum.

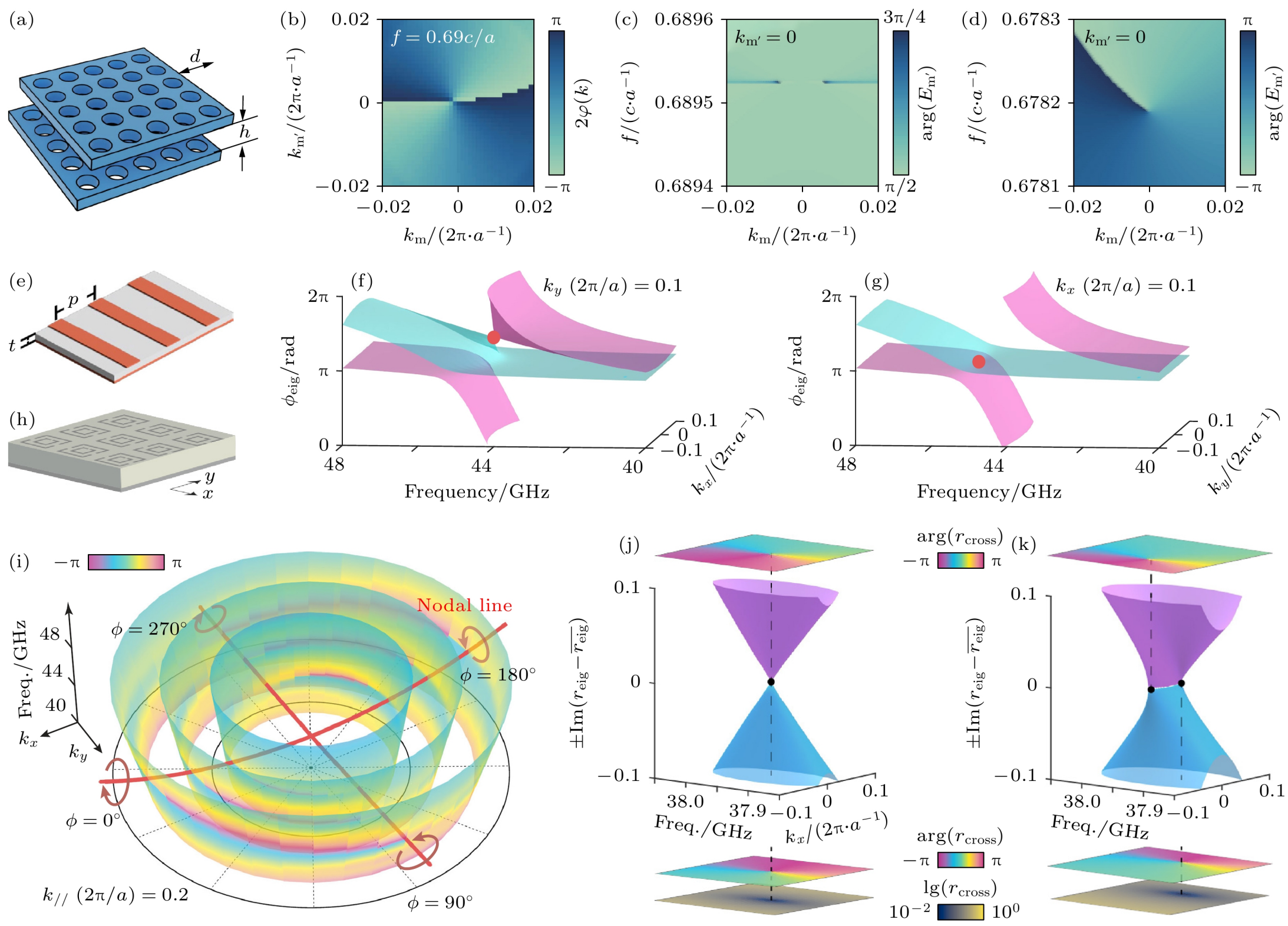

**Fig. 4 Phase fields in momentum-frequency space: (a)–(d) Schematic of a bilayer photonic crystal slab with in-plane offset (a), and the corresponding cross-polarization conversion phase field distributions in momentum space (b) as well as in the momentum-frequency synthetic space (c), (d)[88]; (e) schematic of a one-dimensional reflective photonic crystal slab; (f), (g) phase eigenvalue surfaces of the scattering matrix near a bound state in the continuum in momentum-frequency space[121]; (h) schematic of a printed circuit board-based photonic crystal slab; (i) relationship between the cross-polarization scattering phase vortex and the nodal chain near a bound state in the continuum[121]; (j), (k) introduction of material loss leads to the transformation of diabolic points into exceptional points in the scattering matrix[89].**

The phase field in momentum-frequency space is also significant for understanding the richness of topological phenomena in non-Hermitian systems and their applications[120]. Section 2 introduced the scattering matrix as a reliable tool for describing mode behavior in photonic crystal slabs. Here, we further review the topological singularities inherent in the scattering matrix. Taking a one-dimensional reflective photonic crystal slab as an example (Fig. 4(e)), its scattering matrix can be expressed in the orthogonal basis composed of s-polarization and p-polarization as $\boldsymbol{S}(\omega, k_{//}) = \begin{pmatrix} r_{\mathrm{pp}} & r_{\mathrm{ps}} \\ r_{\mathrm{sp}} & r_{\mathrm{ss}} \end{pmatrix}$. If material loss is neglected, this scattering matrix is unitary. Under the constraints of lattice symmetry, Lorentz reciprocity, and time-reversal symmetry, the scattering matrix can be described by three real numbers $a$, $\phi_1$, and $\phi_2$[121]:

$$\boldsymbol{S}(\omega, k_{//}) = \begin{pmatrix} a\mathrm{e}^{\mathrm{i}(2\phi_2-\phi_1)} & \sqrt{1-a^2}\mathrm{e}^{\mathrm{i}\phi_2} \\ \sqrt{1-a^2}\mathrm{e}^{\mathrm{i}\phi_2} & -a\mathrm{e}^{\mathrm{i}\phi_1} \end{pmatrix}. \quad (26)$$

Research indicates that because the bandwidth of bound states in the continuum (BICs) approaches zero, the eigenphase of their scattering matrix undergoes an abrupt 2π jump at the corresponding frequency. Although this phase jump is topologically trivial in a mathematical sense, it represents a special topological feature absent in Hermitian systems. Figures 4(f) and 4(g) display the phase eigenvalue surfaces of the scattering matrix near the BIC in momentum-frequency space. A conical eigenvalue degeneracy point, known as a diabolic point, is visible. This reveals the anchoring relationship between BICs and diabolic points. Studies further show that encircling this diabolic point once in momentum-frequency space causes the orientation of the scattering matrix eigenpolarization to accumulate a π phase change, reflecting its topological nature. Figure 4(i) presents the cross-polarization scattering phase for the conversion from clockwise to counterclockwise circular polarization. Clear phase vortices and nodal chains are simultaneously observed, further elucidating the intrinsic connection between Hermitian and non-Hermitian systems.

When material loss is considered, the unitarity of the scattering matrix is broken[89]. Taking a photonic crystal slab fabricated on a printed circuit board (Figure 4(h)) as an example: in the lossless case, the diabolic point is locked with the cross-polarization scattering phase corresponding to linear polarization conversion (Figure 4(j)). However, introducing loss transforms the diabolic point into an exceptional point. These exceptional points retain topological robustness in momentum-frequency space. Nevertheless, the positions of their associated cross-polarization scattering phase vortices in frequency-momentum space vary with different incidence planes (Figure 4(k)).

The phase field in momentum-frequency space provides an important pathway for achieving three-dimensional light field control. Currently, fundamental research in this field focuses mainly on topological phenomena in simple systems, while application exploration centers on phase vortices. Future studies need to further reveal the topological properties of scattering matrices in complex non-Hermitian systems. Additionally, they should inversely design and expand the types and functions of phase fields to meet practical application requirements.

### 3.2 Phase Fields in Momentum-Parameter Space

Synthetic space phase fields can also be used to study other radiative topological singularities. Unidirectional guided resonance is an eigenmode that radiates in only one direction. Although it possesses a unidirectional polarization topological charge, it lacks topological robustness in two-dimensional space. Describing its dynamic behavior requires expansion into three-dimensional space. To address this, research proposes using the phase vector field in momentum-parameter space, $(\mathrm{Re}[c^{s}_{x;+,-}/c_0], \mathrm{Im}[c^{s}_{x;+,-}/c_0])$, to mark the position of unidirectional guided resonance and define topological invariants accordingly[84]:

$$S = \frac{1}{2\pi}\oint_C \mathrm{darg}\,[\frac{c^{s}_{x;+}}{c_0}], \qquad (27)$$

The variables in this equation are defined consistently with Eq. (1). The superscript $s =$ t,b distinguishes between upward and downward radiation, which differs fundamentally from the definition of polarization topological charge under bidirectional radiation. Figure 5(a) presents two phase vortices carrying topological invariants of opposite signs in the momentum-parameter space. These topological invariants interact with bulk Fermi arcs (Figures 5(b) and (c)) and symmetry-protected degeneracy points (Figure 5(d)) to enable interband transitions. In anisotropic waveguides, the topological properties of unidirectional guided-mode resonances can be described by phase vortices (phase singularities) in the parameter space[122].

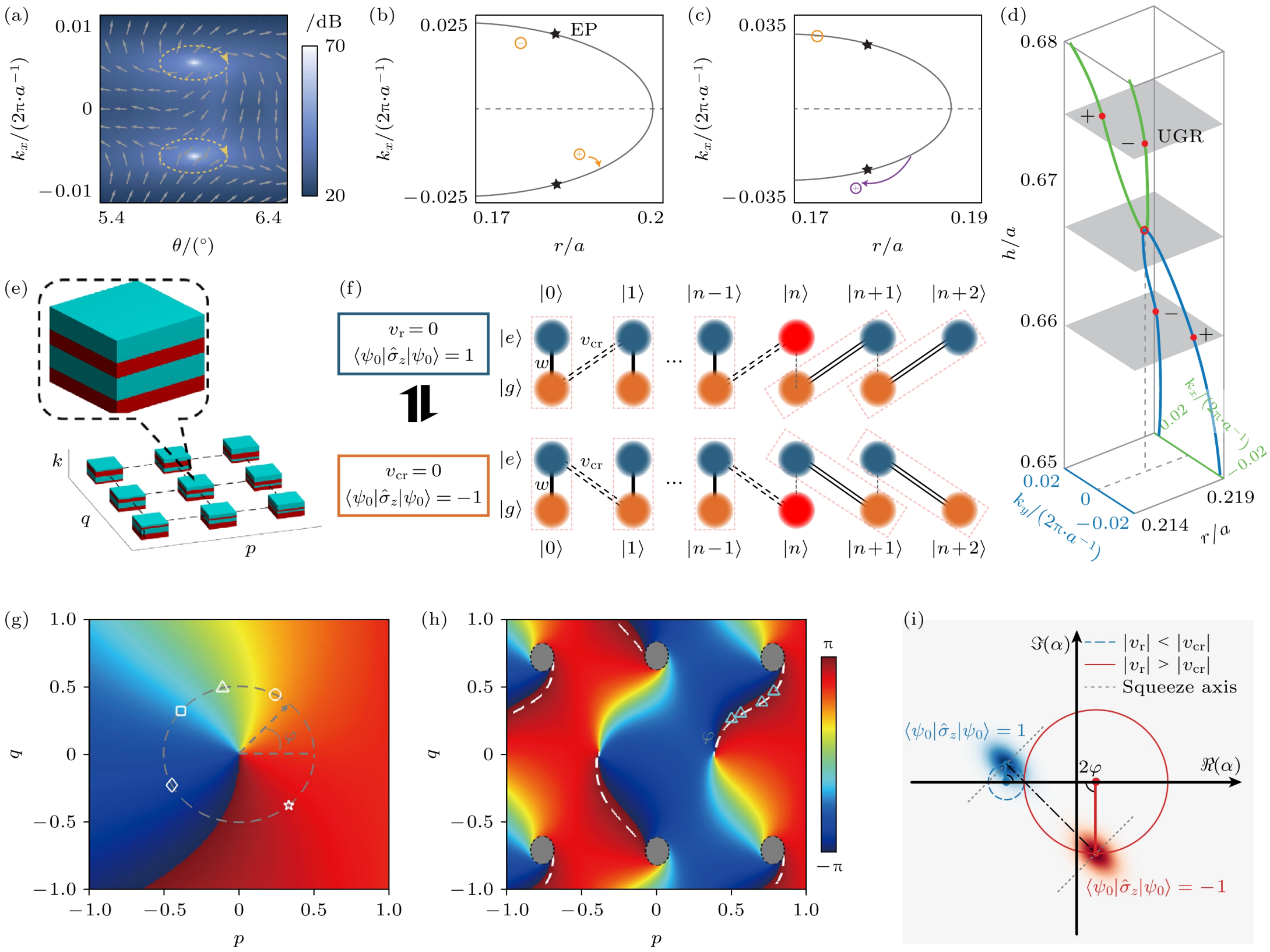

**Fig. 5  Phase fields in parameter-momentum synthetic space: (a) Two phase vortices carrying opposite-sign topological invariants in momentum-parameter space[84]; (b), (c) interaction between phase singularities and bulk Fermi arcs in momentum-parameter space[84]; (d) interaction between phase singularities and symmetry-protected degenerate points in momentum–parameter space[84]; (e) schematic of a two-dimensional array formed by one-dimensional photonic crystals; (f) schematic of a bilayer Fock-state lattice; (g), (h) phase vortices in the parameter synthetic space $(p, q)$[123]; (i) two phase-space circular trajectories[124].**

Furthermore, the phase field within the parameter-synthesized space provides an effective platform for investigating band topology in Hermitian systems. For instance, a two-dimensional array composed of one-dimensional photonic crystals can exhibit linear dispersion relations in its wave vector space. This one-dimensional photonic crystal slab consists of a four-layer structure with $HfO_2$ and $SiO_2$ alternating along the $k$ direction. The thickness of each layer is modulated by parameters $p$ and $q$ (Figure 5(e)). Its Hamiltonian, $H = pv_{pz}\sigma_z + qv_{qy}\sigma_y + \zeta_k v_{kx}\sigma_x$, adopts the typical form of a Weyl Hamiltonian and hosts a Weyl point at $(p, q, k) = (0,0,0.5k_0)$ in the parameter space. When light is incident normally at the frequency corresponding to this Weyl point, the reflection coefficient can be expressed as $r = \mathrm{e}^{\mathrm{i}\phi}$. Studies have revealed that the phase $\phi$ forms a phase vortex[123] in the parameter-synthesized space $(p, q)$, with the vortex center located at the Weyl point (Figure 5(g)). If a reflective substrate is added beneath the structure to introduce an additional phase $\phi_{\mathrm{PC}}$, interface states will be excited when this phase satisfies the condition $\phi_{\mathrm{S}} + \phi_{\mathrm{PC}} = 2m\pi (m \in \mathbb{Z})$ with the intrinsic phase of the structure. The white dashed lines in Figure 5(h) illustrate the distribution trajectory of these interface states in the parameter space. Notably, this trajectory connects multiple reflection phase vortices, clearly revealing the intrinsic connection between interface states and Weyl points.

By leveraging the concept of parameter-synthesized space, synthetic Fock state ($|n\rangle$) lattices can be constructed within single-atom systems. Such systems are described by the quantum Rabi model[124]. Figure 5(f) displays two layers of such lattices. Each lattice layer comprises Fock states and spin states ($|e\rangle$ and $|g\rangle$) connected via intracell coupling $w$ and intercell coupling $v$. The zero-energy defect state is characterized by $\langle\psi_0|\sigma_z|\psi_0\rangle = \mathrm{sgn}(|v_{\mathrm{cr}}| - |v_{\mathrm{r}}|)$. Due to the lack of translational symmetry in the system, traditional real-space band topological invariants are no longer applicable. Consequently, research has shifted toward characterizing these systems using the winding number in phase space[124]:

$$W = \frac{1}{2\pi\mathrm{i}}\int_0^{\pi} \mathrm{d}\varphi \frac{\mathrm{d}}{\mathrm{d}\varphi} \ln \alpha(\varphi) = \frac{1-\mathrm{sgn}(|v_{\mathrm{cr}}|-|v_{\mathrm{r}}|)}{2}, \tag{28}$$

Here, $\alpha(\varphi)$ denotes the circular trajectory in phase space. Figure 5(i) displays two phase-space circular trajectories (red: $|v_r| = 1, |v_{cr}| = 0.25$; blue: $|v_r| = 0.25, |v_{cr}| = 1$), where their intersections correspond to zero-energy defect states. This phase-space representation provides an intuitive approach for distinguishing the fermionic and bosonic characteristics of defect states.

The phase field in the momentum-parameter synthetic space offers a flexible platform for investigating radiative topology and band topology. Looking forward, the research framework for this phase field can be extended to emerging areas in synthetic spaces, such as long-range coupled nonlinear systems, multiphoton simulations, quantum walks, and quantum entanglement.

# 4 Optical State Manipulation in Momentum-Space Phase Fields

## 4.1 Generation of Orbital Angular Momentum

Owing to its topological protection, the momentum-space phase field exhibits robustness against fabrication defects and environmental perturbations. Furthermore, its flexible design principles are not constrained by specific material systems. These advantages highlight its significant application potential in wavefront shaping and quasiparticle manipulation, including orbital angular momentum generation, beam shifts, and half-soliton excitation. This section introduces key applications of momentum-space phase fields in the manipulation of optical states.

Photons possess two fundamental properties: spin angular momentum and orbital angular momentum. Under the paraxial approximation, spin angular momentum is associated with the polarization state of light, manifesting as the rotation of the electric field vector around the propagation direction. In contrast, orbital angular momentum corresponds to the helical distribution of the wave vector around the optical axis, forming a wavefront with a spiral phase. Therefore, these two properties are theoretically independent. However, in spatially inhomogeneous optical fields, they can couple through spin-orbit interaction[100]. In photonic crystal slabs, this interaction enables the conversion of momentum-space phase vortices into real-space spiral wavefronts. Figure 6(d) illustrates the basic method for generating momentum-space geometric phases based on cross-circular polarization conversion, with the physical mechanism detailed in Section 2.2. The generated vortex beams exhibit a ring-shaped distribution (quasi-Bessel beams) in the far field, with a dark center (Figure 6(a)). The carried vortex information can be characterized by interference fringes produced with a linearly polarized reference beam: the number of spiral branches equals the absolute value of the topological charge, while the spiral direction corresponds to the sign of the topological charge[86]. Additionally, vortex information can be characterized by branching features in self-interference fringes (Figure 6(e))[91]. These characterization methods provide a platform for the experimental study and application of phase-space vortex fields.

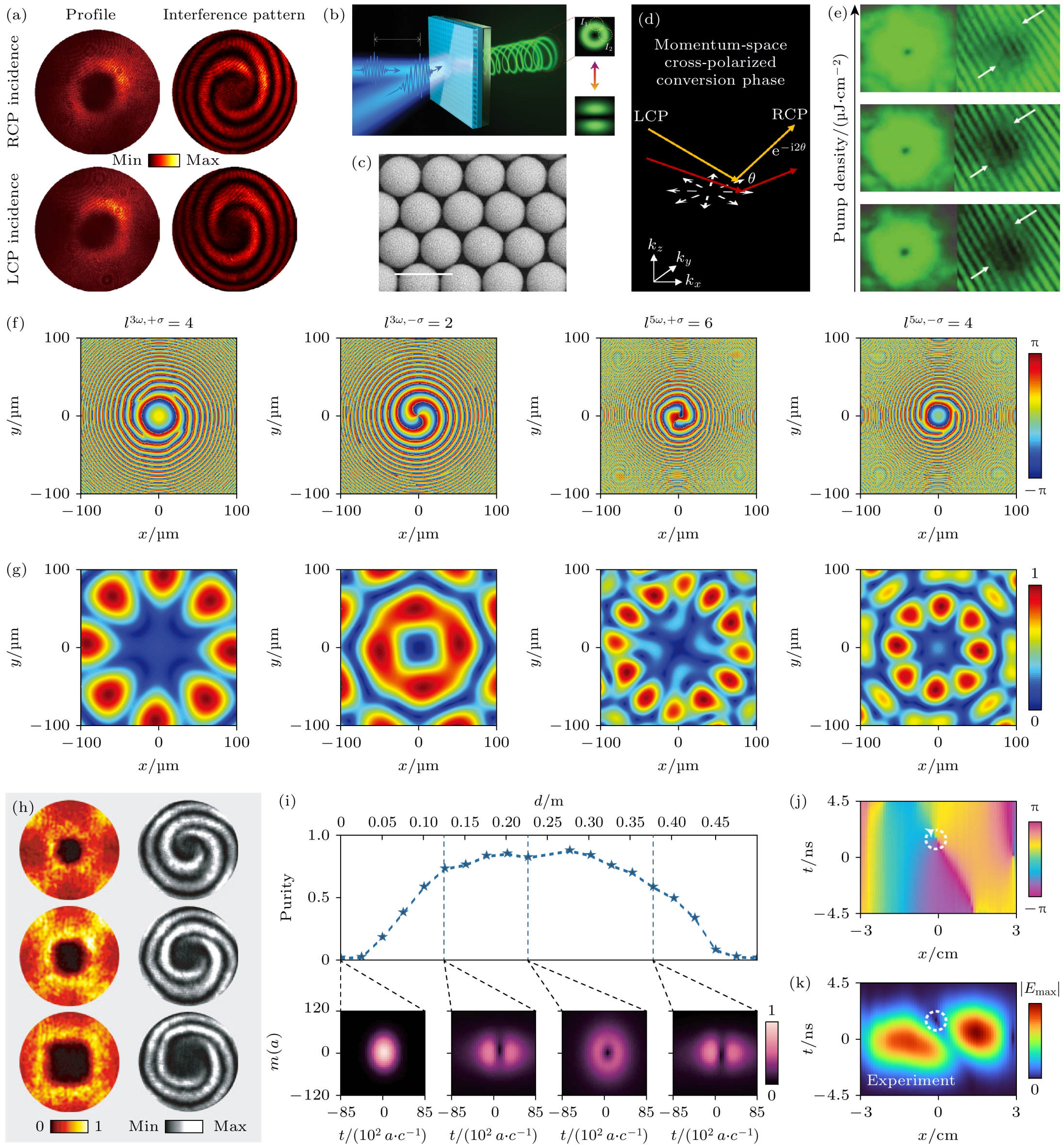

**Fig. 6 Generation of orbital angular momentum: (a) Interference patterns between a vortex beam and a reference beam[86]; (b) ultrafast modulation of a vortex beam[91]; (c) self-assembled photonic lattice structure[125]; (d) schematic of momentum-space geometric phase field formation via cross-polarization conversion; (e) self-interference patterns of a vortex beam[91]; (f), (g) vortex beams generated via high-order harmonic waves[128]; (h) generation efficiency of vortex beams based on cross-polarization conversion[108]; (i) modulation of a spatiotemporal optical vortex[88]; (j), (k) far-field phase and amplitude distributions of the spatiotemporal optical vortex[89].**

The manipulation of vortex beams can be primarily categorized into two types: structural control and optical control. Structural control is achieved by modifying the lattice structure of photonic crystals. For instance, square lattices typically support the generation of vortex beams with a topological charge of $l = \pm 2$, whereas higher-order bound states in the continuum (BICs) in triangular lattices support vortex beams with a topological charge of $l = \pm 4$. Optical control achieves functional switching by adjusting external optical parameters. For example, controlling the operating wavelength allows the topological charge of vortex beams generated by triangular lattice photonic crystal slabs to switch between $l = \pm 4$ and $l = \mp 2$[86]. In perovskite photonic crystal slabs, introducing a spatial offset between two beams forms an elliptical incident spot, which breaks the system symmetry and causes the vortex beam to degenerate into two linearly polarized diffracted beams. Furthermore, introducing a time delay between the two beams (Figure 6(b)) enables ultrafast switching between vortex and linearly polarized light within 1.5 ps[91]. Additionally, nonlinear optical processes allow low-order lattices to generate high-order vortex beams. For the *n*th harmonic, the topological charge $l$ of the vortex beam generated via circular polarization conversion satisfies the relation $l = (n \mp 1)q$ with the polarization topological charge $q$ of the BIC. The "+" sign indicates that the output and input polarization states are orthogonal, while the "$-$" sign indicates they are identical. Figures 6(f) and 6(g) show the results when left-handed circularly polarized light ($\sigma_-$) is incident on an $\alpha$-Si photonic crystal slab ($q = 1$). The generated right-handed circularly polarized ($\sigma_+$) third harmonic carries a topological charge of $l^{3\omega,\sigma_+} = 4$, while the left-handed circularly polarized third harmonic corresponds to a topological charge of $l^{3\omega,\sigma_-} = 2$. For the fifth harmonic, the corresponding topological charges are $l^{5\omega,\sigma_+} = 6$ and $l^{5\omega,\sigma_-} = 4$, respectively. The flexible manipulation of vortex beams based on momentum-space phase vortices offers new possibilities for applications such as high-dimensional optical communication and high-precision optical measurement.

The flexibility in generating vortex beams afforded by momentum-space phase vortices is also reflected in the choice of excitation methods and material systems. Compared to transmission-type schemes, generating vortex beams via reflection-type polarization conversion offers the distinct advantage of tunable efficiency. Studies have found that when material absorption loss is negligible or much smaller than radiation loss, the efficiency of generating vortex beams via reflection-type polarization conversion in a single radiation channel can approach 1. Taking a $Si_3N_4$ photonic crystal slab with a silver mirror substrate as an example, Figure 6(h) shows the corresponding far-field ring distributions from top to bottom for incident left-handed circularly polarized light at wavelengths of 774.9 nm, 777.2 nm, and 779.1 nm. The spot gradually expands outward with changing wavelength while maintaining high conversion efficiency[108]. Figure 6(c) displays an array structure formed by the self-assembly of polystyrene microspheres, which further reduces fabrication costs and provides a solution for the large-scale production of vortex beams based on momentum-space phase vortices[125].

Distinct from the longitudinal orbital angular momentum generated by momentum-space phase vortices, phase vortices in the momentum-frequency synthetic space can be used to generate transverse orbital angular momentum. Research indicates that topological singularities in the scattering matrix of photonic crystal slabs, such as diabolic points and exceptional points, correspond to phase vortices in the momentum-frequency space. Through frequency-domain Fourier transformation, the phase structure of cross-polarization conversion (Figure 6(j)) and its corresponding far-field ring distribution (Figure 6(k)) can be directly observed in the spatiotemporal domain[89]. The purity of spatiotemporal vortices can be characterized by the uniformity of the phase distribution in the spatiotemporal plane (Figure 6(i)). In systems controlled by bilayer lattices[88], adjusting the interlayer horizontal offset allows for the manipulation of the far-field ring distribution morphology while maintaining the spatiotemporal phase vortex. This enables switching between complete and split states (the rightmost to the third-from-right images in the second row of Figure 6(i)), providing a feasible control pathway for optimizing spatiotemporal optical vortices at the micrometer scale. Vortex beams generated by phase vortices in momentum space offer significant advantages due to their compact structure and lack of dependence on a structural center. In recent years, a type of "perfect spatiotemporal vortex" that does not rely on the traditional definition of topological charge has emerged[126,127]. It demonstrates excellent potential in fields such as information processing but currently often relies on complex optical path systems for implementation. In the future, phase field control based on momentum space or momentum-frequency synthetic space is expected to provide more compact, stable, and easily integrable solutions for such novel optical fields.

## 4.2 Optical Topological Spin Textures

Optical spin textures refer to the spatial configuration of light field spin states in real space or momentum space. Currently, significant progress has been made in researching spin structures such as Merons and Skyrmions in real space[129]. However, their realization typically relies on precise optical path design and strict alignment requirements. In contrast, optical spin textures in momentum space can avoid these limitations. When light with specific polarization and incident angle irradiates a photonic crystal slab, it interacts with the non-trivial Berry phase in the system, thereby forming Meron-type spin structures in the momentum space of the output light[130]. Furthermore, if circularly polarized light is incident on a photonic crystal slab supporting BICs, the component of the output light orthogonal to the incident polarization forms a geometric phase vortex field in momentum space, while the co-polarized component does not exhibit this structure. The superposition of these two phase fields directly constitutes the spin texture in momentum space[90]. Figure 7(a) shows the distribution of the curl ($v$) of the Skyrmion texture formed by the phase difference between right-handed and left-handed circularly polarized components in momentum space. This satisfies the relation $v = 2q$ with the topological charge ($q$) carried by the BIC. The polarization of the incident light determines the polarity ($p$) of the texture: right-handed circular polarization corresponds to an upward direction of the central Stokes vector ($p = 1$), while left-handed circular polarization corresponds to a downward direction ($p = -1$). From this, the Skyrmion number corresponding to this momentum-space spin texture can be derived as $N_{\mathrm{sk}} = (1/2)p \cdot v$, where the coefficient $1/2$ represents the ratio of the Meron mapping region to the Poincaré sphere (Figure 7(b)). This equation also indicates that the topological properties of such spin textures can be dynamically controlled and switched via incident polarization. Additionally, their operating wavelength exhibits good scalability, demonstrating broad application potential in optical communication, quantum information processing, and ultra-precise sensing and measurement.

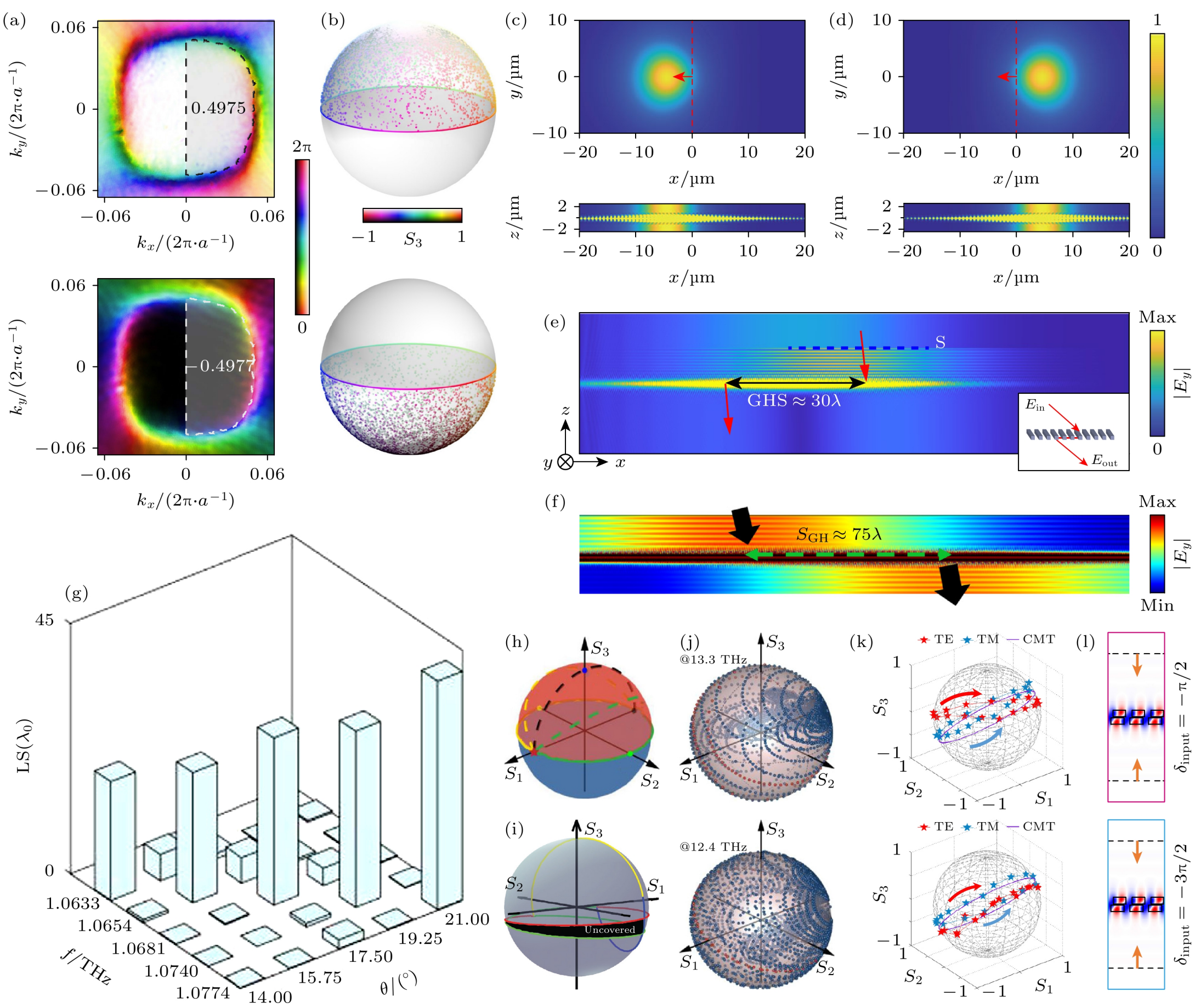

**Fig. 7 Optical state manipulation via momentum-space phase fields: (a), (b) Topological spin textures in momentum space[90]; (c), (d) Goos–Hänchen shift under broken inversion symmetry[110]; (e) Goos–Hänchen shift assisted by a bound state in the continuum under oblique incidence[109]; (f) Goos–Hänchen shift assisted by a unidirectional guided resonance[75];**

**(g) Goos–Hänchen shift assisted by a momentum-mismatched bound state in the continuum[92]; (h) full conversion from arbitrary input polarization to arbitrary output polarization[132]; (i), (j) topological robustness against material loss[112] (i) and spectral scalability[132] (j) of polarization conversion based on momentum-space phase-difference fields; (k) complete polarization conversion based on the critical coupling condition with coherent excitation[133]; (l) perfect coherent absorption assisted by tuning the phase difference between upward and downward radiation[94].**

## 4.3 Beam Shift

Transverse beam shifts are significant in fields such as information storage and ultrasensitive sensing. Traditional transverse shifts, such as the Goos-Hänchen and Imbert-Fedorov shifts, typically rely on oblique incidence conditions. Furthermore, their magnitudes are usually much smaller than the beam waist scale, making direct observation and flexible control difficult. Research indicates that phase gradient fields in momentum space can achieve significantly larger beam shifts. As described in Section 2.3, when linearly polarized light is normally incident on a photonic crystal slab with broken inversion symmetry supporting bound states in the continuum (BICs), a significant geometric phase gradient is generated at the center of the Brillouin zone. Figure 7(c) shows that a Gaussian beam incident in this manner produces a transverse shift of approximately six wavelengths[110]. If oblique incidence is employed near the BIC, a larger phase gradient can be introduced, achieving a shift of approximately 30 wavelengths[109] (Figure 7(e)). Compared to traditional BICs, momentum-mismatched BICs exhibit a broader response range in momentum space[92]. When a Gaussian beam is incident at various oblique angles, significant and easily observable transverse shifts can be achieved (Figure 7(g)). This shift magnitude is proportional to the quality factor $Q$ of the system: $L = v_{\mathrm{g}} Q/(2\pi f_0)$, where $v_{\mathrm{g}}$ is the group velocity. A similar relationship is observed in unidirectional guided-mode resonance: $S_{\mathrm{GH}} = 4Qv/\omega$, where $v$ is the speed of light in the background material[75]. Figure 7(f) demonstrates that when Gaussian light is incident on a photonic crystal slab possessing $P$ symmetry but lacking $C_2^z$ symmetry, a beam shift of approximately 75 wavelengths is observed.

Beyond pursuing large shift magnitudes, controlling the shift direction is also a research focus. Figures 7(c) and 7(d) indicate that when the incident polarization switches from $|+45°\rangle$ to $|-45°\rangle$, the direction of the beam shift reverses[110]. However, the shift occurs only along a single direction in this case. Section 2.3 noted that superimposing the geometric phase field induced by polarization conversion with the cross-polarized resonant phase field can form phase gradients in arbitrary directions within momentum space[111]. Based on this, in-plane oblique beam shifts can be achieved, and the shift direction can be controlled by the spin angular momentum of the incident light. This phenomenon is also known as the photonic spin Hall effect. Future research may further consider introducing orbital angular momentum[131] to enrich the control dimensions and functional characteristics of beam shifts.

### 4.4 Perfect Polarization Conversion and Coherent Absorption

In the radiation control of photonic crystal slabs, phase difference fields are closely related to polarization manipulation and coherent operations. Section 2.4 pointed out that in the phase difference field formed by the phases of the p-polarized and s-polarized components in reflected light, the center of the phase vortex corresponds to perfect polarization conversion between the two[93]. When material absorption loss is neglected, complete conversion from any incident polarization to any output polarization can be achieved near this vortex (Figure 7(h)). Even in the presence of material loss, only a few polarization states remain uncovered (Figure 7(i)), indicating that this mechanism has good stability[112]. Figure 7(j) further shows that this polarization conversion performance is maintained over a wide frequency range[132]. Perfect polarization conversion can also be achieved through the phase difference between two non-coplanar beams. According to the critical coupling condition, the output polarization state is jointly controlled by frequency, radiation coupling strength, and the phase difference between the beams[133]. In systems without material loss, polarization conversion becomes the dominant loss channel, resulting in coherent perfect polarization conversion. If two non-coplanar beams are incident at frequencies near the BIC and system loss is negligible, adjusting the phase difference between the two beams can achieve polarization outputs covering the entire Poincaré sphere (Figure 7(k)). Conversely, when the system does not support polarization conversion, material loss dominates, and the system exhibits coherent perfect absorption. By controlling the phase difference between upward and downward radiation channels, non-coplanar dual beams can be incident with arbitrary relative phases to achieve perfect coherent absorption[94] (Figure 7(l)). In summary, phase difference fields provide key physical guidance and control dimensions for the radiation behavior of photonic crystal slabs.

## 5 Lasers Based on Momentum-Space Phase Fields

Phase fields in momentum space can act as optical resonators and are key elements in constructing micro-nano lasers. Using the polarization conversion phase vortex field formed in perovskite photonic crystal slabs under external circularly polarized optical pumping, low-threshold vortex lasing has been achieved[91] (Figure 8(a)). In addition to directly fabricating micro-nano structures from active materials, gain media can be introduced via spin-coating or covering. For example, by covering a $Si_3N_4$ square-lattice photonic crystal slab with a dielectric layer doped with dye molecules in dimethyl sulfoxide, spin-locked vortex laser output was achieved using the intrinsic phase vortex field of the structure itself[85] (Figure 8(b)). Unlike schemes based on polarization conversion phase vortex fields, this structure has no special requirements for the polarization state of the pump light. Similarly, spin-locked vortex lasers have been successfully realized in $SiN_x$ square-lattice photonic crystal slabs covered with a monolayer of $WS_2$[99] (Figure 8(c)). It is worth noting that in spin-locked vortex lasers, the spiral phase of the wavefront is encoded only within the left- and right-handed circularly polarized components, respectively. Since these components carry topological charges of opposite signs, the overall lasing beam does not carry net orbital angular momentum. Overcoming this limitation is a key scientific problem that needs to be addressed in the future. Furthermore, compared to mature optical pumping schemes, the realization of electrically pumped lasers faces more severe challenges, particularly regarding carrier absorption in the active region. Under electrical injection conditions, precisely analyzing and effectively managing Joule heating is a key pathway to promoting the practical application and on-chip integration of phase vortex field devices. On another front, metasurfaces have made significant progress in optical field phase control. By integrating with dye molecules, lasers with arbitrarily tunable wavefronts can be realized[134] (Figure 8(d)). Recent studies indicate that introducing periodic perturbations in photonic crystal slabs can generate real-space and momentum-space topological singularities[135]. However, the interaction mechanism between momentum-space phase fields and the phase fields constituted by metasurface spatial arrangements remains unclear. Additionally, drawing on metasurface arrangement strategies, arranging different types of photonic crystals in specific ways can achieve lasers with high-order polarization topological charges[136] (Figure 8(e)). Nevertheless, the impact of such macroscopic arrangements on microscopic momentum-space phase evolution remains to be further elucidated.

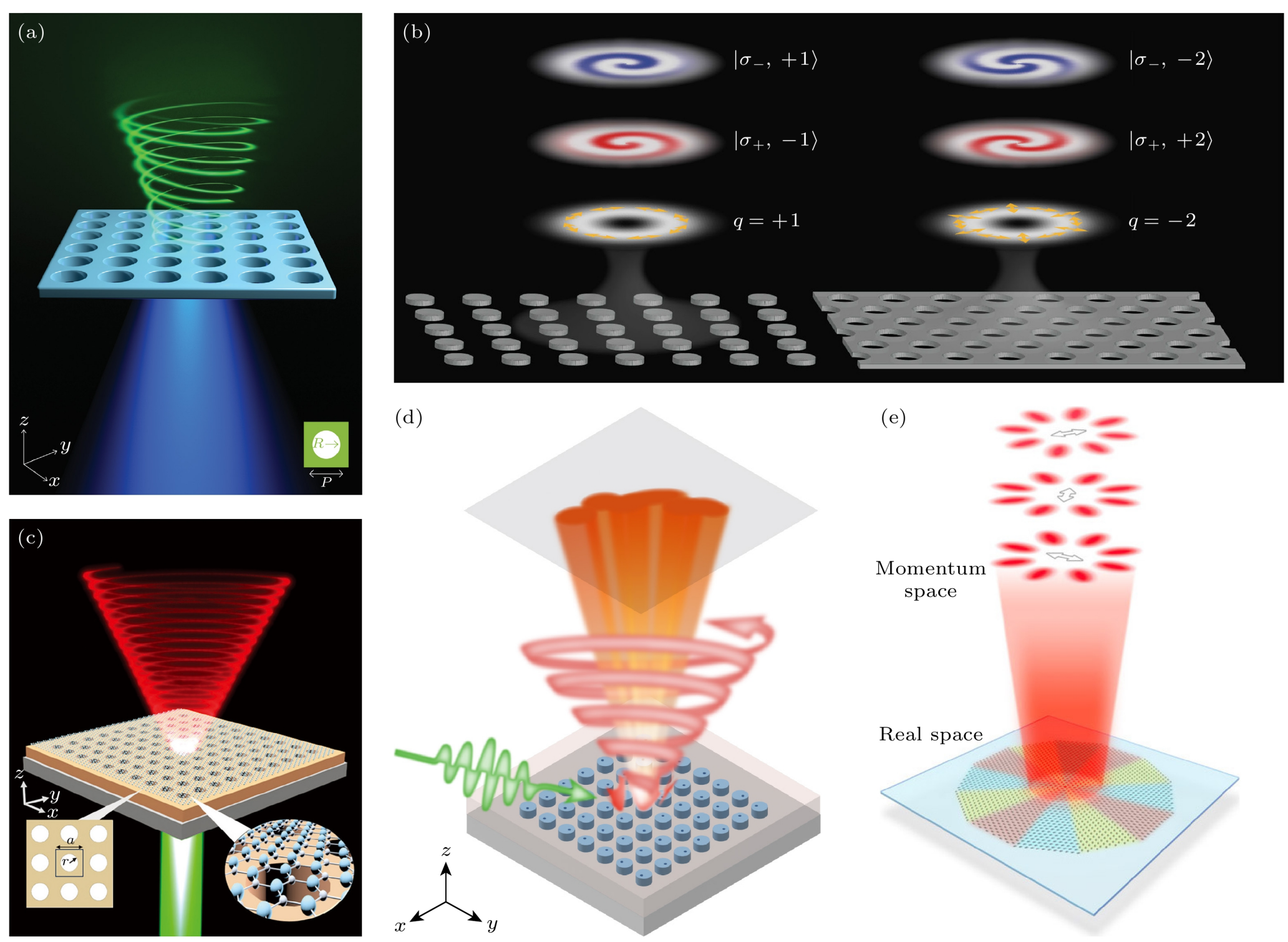

**Fig. 8 Lasers based on momentum-space phase fields: (a) Optically pumped perovskite vortex laser[91]; (b) optically pumped dye-based vortex laser[85]; (c) optically pumped $WS_2$-based vortex laser[99]; (d) optically pumped dye-based laser with arbitrary wavefront generation[134]; (e) optically pumped dye-based laser with tailored topological charge[136].**

# 6 Conclusion and Outlook

By systematically reviewing the momentum-space phase fields of photonic crystal slabs, this article elucidates the pivotal role of phase fields in linking topological properties with optical field modulation functions. This analysis is based on temporal coupled-mode theory and the scattering matrix framework (see Appendix Table A1 for definitions and units of relevant physical quantities). We focus on the generation mechanisms of momentum-space phase vortex fields, phase gradient fields, phase difference fields, and synthetic space phase fields. Furthermore, we discuss their applications in orbital angular momentum generation, the photonic spin Hall effect, momentum-space topological quasiparticles, complete polarization conversion, perfect coherent absorption, and multidimensional optical field modulation. Optical state modulation based on momentum-space phase fields shows broad prospects in integrated photonic devices and advanced optical field control.

Nevertheless, the development of momentum-space phase fields currently faces several key challenges. First, although spin-locked phase vortex fields relax the requirements for orbital angular momentum generation, they remain constrained by spin degeneracy. Observing orbital angular momentum often requires external optical components, lacking on-chip decoupling solutions. Second, electrically pumped excitation is crucial for on-chip photonic integration, particularly for developing high-integration, low-energy multifunctional light sources. Currently, achieving stable excitation of spin-locked orbital angular momentum and momentum-space topological spin structures under electrical injection remains a bottleneck requiring breakthroughs. Finally, introducing inverse design algorithms based on artificial neural networks holds promise for precise prediction and on-demand customization of complex high-order phase fields. Meanwhile, extending the concept of momentum-space phase modulation to quantum information processing and quantum computing will constitute an important direction for future research.

# Appendix A List of Main Symbols

**Table A1** List of principal symbols.

| Symbol | Physical Meaning | Unit |
|---|---|---|
| cx, cy | Polarization vector | Modulus: V/m; Phase: rad |
| dx, dy | Radiative coupling coefficient (x, y polarization basis) | Modulus: s-1/2; Phase: rad |
| ds, dp | Radiative coupling coefficient (s, p polarization basis) | Modulus: s-1/2; Phase: rad |
| E | Electric field vector | Modulus: V/m; Phase: rad |
| γ0 | Radiative decay rate | rad/s |
| k | Wave vector | m-1 |
| l | Momentum-space phase winding number | Dimensionless |
| r | Reflectance | Dimensionless |
| R | Reflection coefficient | Modulus: Dimensionless; Phase: rad |
| σ-, σ+ | Left/right-handed circular polarization basis | Dimensionless |
| S0-S3 | Stokes parameters (normalized) | Dimensionless |
| t | Transmittance | Dimensionless |
| ω | Resonant angular frequency | rad/s |
| ω0 | Resonant center angular frequency | rad/s |